\documentclass[10pt,journal,compsoc]{IEEEtran}
\ifCLASSOPTIONcompsoc
  \usepackage[nocompress]{cite}
\else
  \usepackage{cite}
\fi

\usepackage{tabularx}
\usepackage{graphicx,epsfig,color}
\usepackage{amssymb}
\usepackage{graphicx}
\usepackage{url}
\usepackage{algorithm}
\usepackage{algorithmic}
\usepackage{amsmath}
\usepackage{balance}
\usepackage{float}
\usepackage{comment}
\usepackage{cite}
\usepackage{tikz}
\usepackage{tikz-network}
\usepackage{pgfplots}
\usepackage{subcaption}
\usepackage{multirow}
\usepackage{caption}
\usepackage{chronosys}
\usepackage{xcolor}

\usetikzlibrary{shapes.geometric,positioning,shapes,arrows,calc,automata}

\begin{document}
\title{Green-PoW: An Energy-Efficient Blockchain Proof-of-Work Consensus Algorithm }

\author{Noureddine~Lasla,~\IEEEmembership{Member,~IEEE,}
        Lina~Salim~Alsahan, ~Mohamed~Abdallah,~\IEEEmembership{Senior~Member,~IEEE}
        and~Mohamed~Younis,~\IEEEmembership{Senior~Member,~IEEE}
\IEEEcompsocitemizethanks{\IEEEcompsocthanksitem N. Lasla, L. S. Alsahan and M. Abdallah are with the Division of Information and Computing Technology, College of Science and Engineering, Hamad Bin Khalifa University (HBKU), Doha, Qatar.\protect\\
E-mail: \{nlasla, lialsahan, mabdallah\}@hbku.edu.qa
\IEEEcompsocthanksitem Mohamed Younis is with University of Maryland, Baltimore County, Baltimore, MD, USA.\protect\\
Email:\ younis@umbc.edu
}
\thanks{Manuscript received ; revised .}}

\IEEEtitleabstractindextext{
\begin{abstract}
This paper opts to mitigate the energy-inefficiency of the Blockchain Proof-of-Work (PoW) consensus algorithm by rationally repurposing the power spent during the mining process. The original PoW mining scheme is designed to consider one block at a time and assign a reward to the first place winner of a computation race. To reduce the mining-related energy consumption, we propose to compensate the computation effort of the runner(s)-up of a mining round, by granting them exclusivity of solving the upcoming block in the next round. This will considerably reduce the number of competing nodes in the next round and consequently, the consumed energy. Our proposed scheme divides time into epochs, where each comprises two mining rounds; in the first one, all network nodes can participate in the mining process, whereas in the second round only runners-up can take part. Thus, the overall mining energy consumption can be reduced to nearly $50\%$. To the best of our knowledge, our proposed scheme is the first to considerably improve the energy consumption of the original PoW algorithm. Our analysis demonstrates the effectiveness of our scheme in reducing energy consumption, the probability of fork occurrences, the level of mining centralization presented in the original PoW algorithm, and the effect of transaction censorship attack.   
\end{abstract}

\begin{IEEEkeywords}
Blockchain, Consensus algorithm, Proof-of-Work, Energy-efficiency.
\end{IEEEkeywords}}
\maketitle

\section{Introduction}
Nakamoto's Blockchain protocol \cite{bitcoin}, also known as proof of work (PoW), is the first to achieve consensus in a permission-less setting, where anyone can join or leave during the protocol execution. Its main security design goal is to prevent \textit{Sybil attacks} by relying on a computational cryptographic puzzle-solving process. The protocol has proven its robustness since its first application (Bitcoin) in 2009. However, besides being the most trusted and secure public consensus algorithm, PoW is considered as a computation-intensive voting-based consensus process. For instance, PoW-powered Bitcoin mining consumes massive power that could suffice for a small country, like Denmark \cite{deetman2017bitcoin}. Moreover, as estimated by Digiconomist \cite{mora2018bitcoin}, Bitcoin usage emits about 33.5 MtCO2e annually, as of May 2018. The main reason behind the PoW excessive energy is to make attacks against the blockchain network very expensive. 

In a nutshell, PoW is a leader election protocol that designates among network participants (miners) one leader that will append the next block to the chain. To attract more participants to join and maintain the network, and at the same time demotivate them from cheating, an honest miner can be elected to receive a very attractive reward if it can solve a computationally arduous puzzle. 
The idea behind making a difficult puzzle in PoW consensus algorithm is to bound the economic capacity of an adversary to successfully undermine the network, for instance, to prevent double spending attack and rewriting the block-history \cite{isra, li2020survey}. Bounding the capacity of users based on their computation (energy) is only a \textit{sufficient} condition to prevent security attacks and not \textit{necessary}. In the literature, there have been several attempts to rationalize the consumed energy in PoW by either bounding the miner economic capacity using alternative more energy-efficient mechanisms \cite{badertscher2018ouroboros, PoET, PoR}, or by recycling the wasted energy spent in solving the puzzle to also serve for other useful tasks \cite{REM}. 
However, such rationalization still cannot meet the same security level as the original PoW. Proof-of-Stake (PoS) \cite{badertscher2018ouroboros, bentov2016cryptocurrencies}, for instance, is a  greener  form of distributed consensus where validators (akin to miners) do not have to use their computation power but only proof the ownership of an amount of stake (bond) in order to vote on new blocks. Such an approach suffers from a Nothing-at-Stake problem, where validators can stake for different blocks supporting multiple forks in order to maximize their chances of winning a reward \cite{baliga2017understanding}. Some advanced PoS implementations to overcome the Nothing-at-Stake problem, such as Casper \cite{Ethereum}, are still in the testing stage and not yet deployed in large scale networks. 

In this paper, we propose a new consensus algorithm that rationalizes the computational overhead of the original PoW and reduces its overall energy consumption to nearly $50 \%$ without degrading its security level.  The main idea is to factor the power spent during one mining round, in not only electing which node will write the current mining block but also selecting a small subset of miners $\mathcal{M}'$ that will be allotted exclusivity to mine in the next round (power-save round). Therefore, we modify the original mining scheme by defining a new participation rule in a mining race. Unlike the original PoW, the race in the next round will be among only a small subset of miners $\mathcal{M}'$, and thus less energy will be consumed. The other miners keep waiting until the block gets created before resuming again the mining process. Because only a small subset of miners will participate in a power-save round, it is possible that a deadlock happens if the involved $|\mathcal{M}'|$ miners fail to generate the block. This may occur, for instance, if the miners in $\mathcal{M}'$ get disconnected from the network. To ensure \textit{liveness} and avoid such a situation, we introduce a time-out after each normal race round. If the nodes that are not participating in the mining during a power-save round, do not receive a new block after a specific period of time, they automatically resume the mining in order to create the missing block.

The contributions of this paper can be summarized as follow:
\begin{itemize}
    \item In order to reduce the energy consumption in PoW, Green-PoW considers mining rounds in pairs and adjusts the election mechanism during the first mining round in the pair, to allow the election of a small subset of miners that will exclusively mine in the second round. Assuming the total network hashing power is equally distributed among miners, a significant energy saving up to $50\%$ could thus be achieved.
    \item Given that only a small subset of miners can participate in the even-numbered mining rounds, Green-PoW reduces the probability of fork occurrences. Therefore, the total fork occurrence will be considerably decreased compared to the original PoW algorithm.  
    \item Green-PoW can help in reducing the level of mining centralization presented in PoW. This kind of centralization can happen when a small portion of the miners, e.g., mining pools in the case of Bitcoin, holds the majority of the network hash-rate. In Green-PoW, a winner during the first round cannot participate in the second round mining, thus miners that hold the majority of the network hash-rate will no longer have the same potential for dominating block generation. 
    \item We demonstrate the effectiveness of Green-PoW algorithm in reducing the overall energy consumption, during mining, by conducting a stochastic analysis where the mining is modeled as a Poisson process.
\end{itemize}

The remaining of this paper is organized as follows. Section \ref{sec:related-work} reviews the existing work to improve the energy consumption in PoW. Preliminaries about mining in PoW are presented in Section \ref{sec:preliminaries}. The detailed description of our proposed consensus algorithm can be found in  Section  \ref{sec:green-pow}. Section  \ref{sec:security_analysis} analyzes the security properties and reports the performance results of  Green-PoW.  Finally  the  paper  is concluded in Section \ref{sec:conclusion}.

\section{Related Work}
\label{sec:related-work}
In this section, we survey studies made so far in order to mitigate the energy-inefficiency of the PoW algorithm. There are two main categories of solutions. In the first category, the goal is to improve the energy usage of the original PoW by either recycling the power spent during the mining process in serving other useful real problems rather than solving a useless puzzle or modifying the consensus protocol flow while maintaining the cryptographic puzzle element unchanged. On the other hand, the second category follows completely different and consensus algorithms such as Proof of Stake (PoS) \cite{badertscher2018ouroboros}, Proof of Elapsed Time (PoET)\cite{PoET}, and Proof of Retrievability (PoR)\cite{PoR}. While this class of solutions can achieve considerable energy saving, yet it cannot reach the same security level as the well tested PoW. Our proposed Green-PoW consensus mechanism can be classified in the first category of solutions, which aims to improve the energy consumption of  PoW.  Therefore, the focus in this section is on setting Green-PoW apart from competing in such a category.

REM \cite{REM} replaces the wasted computations in performing the conventional PoW by executing useful workloads outsourced by clients. REM relies on the availability of Software Guard Extensions (SGX) technology at the miners' infrastructure, where a trusted execution environment is utilized to preserve the integrity and confidentiality of the workload. Miners get rewarded based on the number of instructions they can process successfully. However, REM falls short in overcoming the centralized nature of the SGX attestation process where accessing Intel's server is mandatory in order to validate clients' workloads. Unlike REM, A. Shoker \cite{eXercise} and S. King \cite{prime} have proposed two distinct schemes that invest the mining effort to solve public problems that do not require confidentiality. Examples of the considered public problems include the generation of large prime numbers and performing complex matrix operations. In both schemes, the difficulty of solving a public problem can be adjusted, which allows it to be an appropriate replacement for the hashing processes done by the traditional PoW. However, the aforementioned schemes cover only a predefined set of problems which limits the utility of the processed workload.

Following a similar approach, Felipe et al. \cite{PoL} have presented a proof-of-learning consensus mechanism that substitutes the PoW puzzle by machine learning tasks. The goal is to re-purpose the wasted computational power to build a decentralized system for crowd-sourcing machine learning models. In their approach, nodes are classified into suppliers, trainers, and validators. Supplier nodes publish machine learning tasks and provide the training and testing data, whereas trainers perform the training work in an attempt to win the block. Validator nodes, which are chosen randomly, are responsible for ranking the published machine learning models based on their performance, then pick the best as proof of the work. By design, the models, testing, and training data are all public and the scheme does not provide any confidentiality guarantees, which makes the usability of such an approach questionable. In order to preserve the privacy of both training and testing data, Qu et al. \cite{PoFL} have considered the use of federated learning. This enabled crowd-sourcing the workload required to train a model while using miner's local training data. Furthermore, the verification of the model's accuracy is done using a Homomorphic encryption-based verification scheme to prevent the disclosure of testing data to the network.   

Daian et al. \cite{Piece} have proposed outsourcing PoW to defeat several frequent cyberattacks such as spam emails and DDoS to recycle the inevitable wasted power. This can be achieved by utilizing the difficulty of solving the cryptographic problem as a challenge for the clients in order to benefit from a service. The mining process is decomposed into two parts, namely, inner and outer puzzles. The inner puzzles are solved by the service clients named workers that provide solutions to the outsourcer to be verified. The outer puzzle process the solutions provided by the workers in order to find the overall solution of the PoW. Consequently, the cost of solving the inner puzzle controls the number of requests that can be submitted by the client in a short period of time. Implementing outsourced PoW in practice is restricted by the resources used at the client's infrastructure (e.g. PCs, mobile phones). As lightweight clients will not be capable of solving the puzzle as fast as other power-full machines, this may delay or prevent them from reaching the service. Such a concern hinders the fairness of the network and obstructs the reliability of the service. 

One prominent work that modified the consensus protocol to optimize PoW is Bitcoin-NG \cite{bitcoin-ng}. Bitcoin-NG altered the conventional PoW consensus by segregating the election of a block creator from processing transactions. This necessitates dividing Bitcoin's block structure into two new types, key blocks for electing a leader for the next epoch, and micro blocks generated by the epoch's leader and consist of network's transactions. This change in design led to faster transaction validation as micro blocks do not require including proof of work, alongside to faster key block propagation due to its small size, which consequently reduced the wasted energy caused by chain forks. However, Bitcoin-NG suffers from frequent micro block forks that occur whenever a new leader is elected for the new epoch because the current leader will carry on generating micro blocks and propagate it to the network while being unaware that another miner already solved a new key block and started generating micro blocks simultaneously. This behavior makes the network susceptible to fork-related attacks, e.g., double-spending. To mitigate this issue, the authors have proposed a new type of transaction that can be issued by nodes that witnessed two conflicting transactions to report an attempt of double spending and motivate the investigation of the fraudulent miner rewarding. Yet, if the double-spending attack is not noticed during the maturity window of the attacker key block or before the attacker spent the revenue, the double spending will indeed occur and the involved leader will not be penalized. Based on the aforementioned concern we can conclude that although Bitcoin-NG is an enhanced version of Bitcoin in terms of transaction throughput and energy optimization, it has diminished the level of security offered by the original Bitcoin protocol.

In summary, the surveyed studies fall short in effectively addressing the power consumption of PoW. Existing approaches either replace the crypto-puzzle with different types of useful work which adds complexity to the consensus process or alter the ledger's structure and consensus flow drastically, which degrades the network's security. Green-PoW opts to avoid these shortcomings by achieving dramatic energy reduction while sustaining the security properties of the PoW based consensus methodology.

\section{Preliminaries}
\label{sec:preliminaries}
In this section, we present the two main core components of Bitcoin blockchain, namely, the block mining process and difficulty adjustment. In addition, we discuss the security concerns about the fundamental design. 

\subsection{Proof-of-Work}
\subsubsection{Block mining}
In Bitcoin, miners participate in a PoW based consensus in order to maintain the consistency and integrity of a public distributed ledger, which is an ever-growing chain of a tamper-proof data structure called blocks. Each block records a list of transactions, previous block hash, Merkle root, timestamp, hash target, and a nonce. Blocks are chained by storing the hash of the predecessor block in the current block header. Miners participate in an incentivized race to forge a new valid block, by following a brute force approach to find a 32-bit \emph{nonce} value that yields a block hash less than the target, which is derived from an adjustable value called mining difficulty, used to maintain the equilibrium between the block generation rate and the invested computational power over-time\cite{bitcoin} (more details in section \ref{sec:mining-difficulty}).

The mining process in the original PoW is illustrated in \figurename ~\ref{fig:original_bitcoin}. When the mining race begins, miners start competing for forming a valid block; the first miner that finds the nonce is considered as the leader of the current round for creating the new block. Such a miner announces the block to the rest of the network to get rewarded with a newly generated Bitcoins. Every other miner that receives the new block generated by the winning miner immediately desists mining the current block and start mining the next one. 
\pgfdeclarelayer{fg} 
\pgfsetlayers{main,fg} 
\usetikzlibrary{decorations.text}
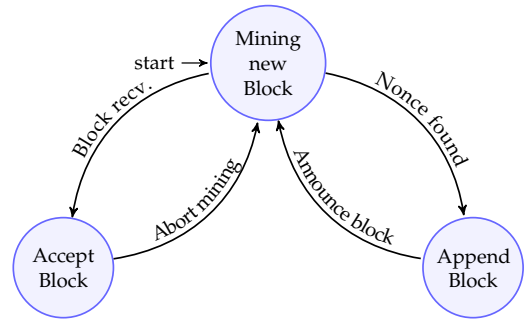
\begin{figure}
    \centering
    \tikzstyle{every state}=[draw=blue!60, fill=blue!5, thick, text width=1.2cm, align=center]
    \scalebox{0.8}{
    \begin{tikzpicture}[->,>=stealth',shorten >=1pt,auto,node distance=2.8cm,semithick]
        \node[state, initial] (A)            {Mining new Block};
        \node[state] (B) [below right=3cm of A] { Append Block};
        \node[state] (C) [below left=3cm  of A]   {Accept Block};
        \draw [->,thick, postaction={decorate,decoration={text along path,text align=center,raise=0.1cm, text={Nonce found}}}]      (A) to [bend left=35] (B);
        
        \draw [<-,thick, postaction={decorate,decoration={text along path,text align=center,raise=0.1cm, text={Block recv.}}}]      (C) to [bend left=35] (A);
        
        \draw [<-,thick,postaction={decorate,decoration={text along path,text align=center,raise=0.1cm, text={Announce block}}}]      (A) to [bend right=35] (B);
        \draw [->,thick,postaction={decorate,decoration={text along path,text align=center,raise=0.1cm, text={Abort mining}}}]      (C) to [bend right=35] (A);

    \end{tikzpicture}
    }
    \caption{The state diagram of a miner in the original PoW.}
    \label{fig:original_bitcoin}
\end{figure}

\subsubsection{Mining difficulty}
\label{sec:mining-difficulty}
Mining difficulty \emph{(D)} is a measure of how difficult it is to find a \emph{Nonce} of a valid block. It is reevaluated every two weeks based on the average block time of the previous 2016 blocks, to maintain a fixed block generation time (10 minutes in Bitcoin)\cite{bitcoin}. The difficulty will increase when the average block time is less than the expected, as it indicates that the network's computational power has increased and miners have become capable of generating new blocks in less than 10 minutes. Sometimes the network experiences a plummet in the total computational power when a set of miners with a significant hashing capability depart the network thus, the mining difficulty will decrease.
Equation (\ref{eq:1}) and (\ref{eq:2}) shows the relation between the previous average block time and the difficulty level, where \emph{F} is the factor used to recalculate the difficulty of the next two weeks. It represents the ratio of the expected average block time of the network \emph{$T_{E}$} to the actual recorded average block time \emph{$T_{Avg}$}. 

Based on how the difficulty adjustment works, Bitcoin pushes miners to invest in more powerful mining rigs, to be qualified for competing with other miners under the incriminating mining difficulty. Despite the operation cost, miners continue to spend more because of the expected remuneration when winning the mining race. This feature makes PoW based blockchain a power-hungry system that burns energy indefinitely as long as the network of miners is growing.  

\begin{equation}\label{eq:1}
F = \dfrac{T_{E}}{T_{Avg}} 
\end{equation}

\begin{equation}\label{eq:2}
D_{(i+1)} = D_{(i)}\times F
\end{equation}
 
\subsubsection{Chain forks}
\label{sec:fork}
Blockchain may witness inconsistencies that cause branching in the chain, this phenomenon is called forks, where more than one valid block is broadcasted to the network simultaneously\cite{forks1}. Consequently, nodes get confused when updating the chain and each will end up adopting different blocks as the chain head. Network propagation delays are mostly the reason for fork occurrence, which has been validated empirically in \cite{LocalSim, forks1}. Other factors also play a role in causing forks such as the time needed to generate a block, network bandwidth, and the number of connections each node establishes with the network. Blockchain nodes resolve a fork in the next block period by adopting the longest chain, which is identified by calculating the cumulative number of expected hashes performed to generate each block in the chain. The frequent occurrence of forks allows adversaries to conduct various attacks (discussed further in section \ref{sec:security_analysis}) that could degrade the security of the blockchain network.

\section{Energy-Efficient Consensus Algorithm}
\label{sec:green-pow}
\tikzstyle{miner}=[draw=red!60, fill=red!5, thick, align=center, circle]
\tikzstyle{winner}=[draw=blue!60, fill=blue!5, thick, align=center, circle]
\tikzstyle{winner2}=[draw=green!60, fill=green!5, thick, align=center, circle]
\tikzstyle{secondW}=[dashed, draw=blue!60, fill=blue!10!red!2, thick, align=center, circle]
\tikzstyle{idle}=[draw=gray!60, fill=gray!5, thick, align=center, circle]
\tikzstyle{block}=[scale=0.5, draw=black, fill=black!2, very thick, scale=2.5, align=center]
\tikzstyle{newblock}=[scale=0.5, draw=blue!60, fill=blue!5, very thick, align=center, scale=2.5]
\tikzstyle{nextblock}=[scale=0.5, draw=green!60, fill=green!5, very thick, align=center,scale=2.5]

\begin{figure*}
\begin{subfigure}{.49\textwidth}
    \centering
    \begin{tikzpicture}
        \node[miner] at (0, 0) {};
        \node[miner] at (0, 1) {};
        \node[miner] at (0.9, 0) {};
        \node[miner] at (-0.8, 0.1) {};
        \node[miner] at (0.6, -0.66) {};
        \node[miner] at (0.2, -1) {};
        \node[miner] at (1.7, -0.3) {};
        \node[miner] at (2.3, -0.3) {};
        \node[winner] at (2.5, 0.5) {};
        \node[miner] at (-1.8, 0.9) {};
        \node[miner] at (-1.4, .0) {};
        \node[miner] at (1.3, -1.2) {};
        \node[miner] at (-.5, -1.2) {};
        \node[miner] at (-2.1, -.2) {};
        \node[miner] at (1.5, .4) {};
        \node[miner] at (-1.5, -0.9) {};
        \node[secondW] at (1.5, 1) {};
        \node[secondW] at (.8, .8) {};
        \node[secondW] at (-1.2, .8) {};
        \node[secondW] at (-.8, -0.6) {};
        
        \node[block] (b0) at (-2,-2.3){$0$};
        \node[block] (b1) at (-1,-2.3){$1$};
        \node[block] (b2) at (0,-2.3){$2$};
        \node[block] (b3) at (1.,-2.3){$3$};
        \node[newblock] (b4) at (2,-2.3){$4$};
        \node[white] (b5) at (3,-2.3){};
        \node[] (b00) at (-3.3,-2.3){};
        \node[] at (4,-2.3){};
        \path (b0) edge  [] node {} (b1);
        \path (b1) edge  [] node {} (b2);
        \path (b2) edge  [] node {} (b3);
        \path (b3) edge  [] node {} (b4);
    \end{tikzpicture}
    \caption{First mining round $\rho_2^1$.}
     \label{fig1:blocks_5miners_1000}
 \end{subfigure}
 \hfill
 \begin{subfigure}{.49\textwidth}
 \centering
    \begin{tikzpicture}
        \node[idle] at (0, 0) {};
        \node[idle] at (0, 1) {};
        \node[idle] at (0.9, 0) {};
        \node[idle] at (-0.8, 0.1) {};
        \node[idle] at (0.6, -0.66) {};
        \node[idle] at (0.2, -1) {};
        \node[idle] at (1.7, -0.3) {};
        \node[idle] at (2.3, -0.3) {};
        \node[idle] at (2.5, 0.5) {};
        \node[idle] at (-1.8, 0.9) {};
        \node[idle] at (-1.4, .0) {};
        \node[idle] at (1.3, -1.2) {};
        \node[idle] at (-.5, -1.2) {};
        \node[idle] at (-2.1, -.2) {};
        \node[idle] at (1.5, .4) {};
        \node[idle] at (-1.5, -0.9) {};
        \node[miner] at (1.5, 1) {};
        \node[miner] at (.8, 0.8) {};
        \node[miner] at (-1.2, .8) {};
        \node[winner2] at (-.8, -0.6) {};
        
        \node[block] (b0) at (-2,-2.3){$0$};
        \node[block] (b1) at (-1,-2.3){$1$};
        \node[block] (b2) at (0,-2.3){$2$};
        \node[block] (b3) at (1.,-2.3){$3$};
        \node[newblock] (b4) at (2,-2.3){$4$};
        \node[nextblock] (b5) at (3,-2.3){$5$};
        \node[] (b00) at (-3.3,-2.3){};
        \node[] at (4,-2.3){};
        \path (b0) edge  [] node {} (b1);
        \path (b1) edge  [] node {} (b2);
        \path (b2) edge  [] node {} (b3);
        \path (b3) edge  [] node {} (b4);
        \path (b4) edge  [] node {} (b5);
    \end{tikzpicture}
    \caption{Second mining round $\rho_2^2$.}
     \label{fig:blocks_5miners_1000}
 \end{subfigure}
\caption{Illustrating the mining process of Green-PoW in one epoch of time ($\tau_2$). The blue node is the winner of the first round ($\omega_2$) and generates the block number $4$ (blue square). Red nodes are losing participants in the mining. The dashed blue nodes are second place winners ($\mathcal{M}_2^2$), which will be the only nodes to handle the block in the next round ($\rho_2^2$). The green node is the winner of the second round which generates block number $5$ (green square). Gray nodes are idle miners that do not participate in the mining during the second round \{$\mathcal{M} \setminus \mathcal{M}_2^2$\}. }
 \label{fig:epoch}
\end{figure*}
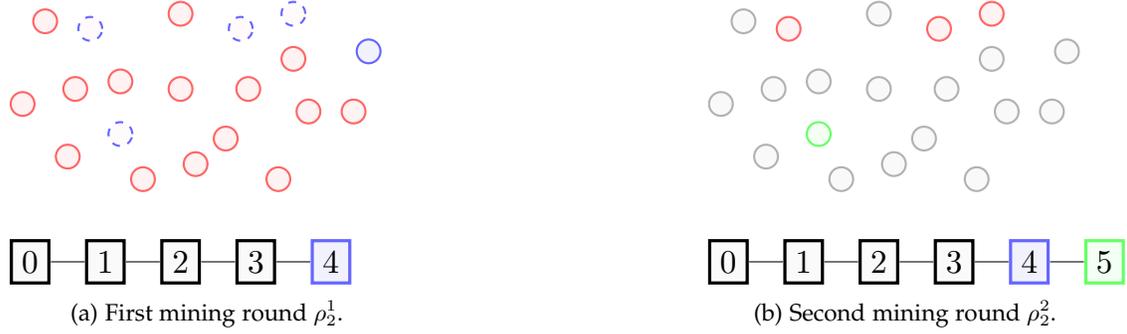
Green-PoW is an energy-efficient consensus algorithm that reduces the computation load to nearly 50\% compared to the original Bitcoin's PoW algorithm, without affecting the other properties of the system. The algorithm divides time into epochs, where each epoch consists of two consecutive mining rounds. Let $\tau_i$ denote the epoch of time corresponding to the creation of blocks number $2i$ and $2i+1$, and $\rho_{i}^1$ and $\rho_{i}^2$ denote the first and second mining round, respectively, within the epoch $\tau_i$. During $\rho_{i}^1$, the mining process to create a new block follows the same Bitcoin mining steps, where the set of all miners, which is denoted as $\mathcal{M}$, can participate. In addition, a very small subset of miners, denoted as $\mathcal{M}_i^2$, is elected during this same mining round to be the only eligible participants for mining the next block in $\rho_{i}^2$. Note that $\mathcal{M}_i^1$, which refers to the set of miners that can participate in $\rho_{i}^1$, is equal to $\mathcal{M}$ and both are used interchangeably. All the other miners $\mathcal{M}\setminus \mathcal{M}_i^2$, during $\rho_{i}^2$, pause until the considered block gets appended before starting a new mining epoch ($\tau_{i+1}$). An illustrative example of the mining process during one epoch of time is given in \figurename \ \ref{fig:epoch}. The detailed description of each step of the algorithm is given in the remaining of this section.

\begin{algorithm}[H]
\caption{Green-PoW Consensus Algorithm.}
\label{alg:greenPoW}
 \begin{algorithmic}[1]
 \renewcommand{\algorithmicrequire}{\textbf{Input:}}
 \renewcommand{\algorithmicensure}{\textbf{Output:}}
  \STATE $init: $\\ $b=1$; $\rho= 1$ \COMMENT {$b:$ block \ number, $b:$ $\rho:$ round}
  \\ $runnerUp= false$ \COMMENT {I am a runner-up}
  \LOOP 
  \IF{$\rho == 1$}
    \STATE $nonce = $FindBlockNonce$(b)$ 
    \IF {$nonce$} 
        \STATE \COMMENT{nonce found}
        \STATE AppendNewBlock$(b)$
        \STATE AnnounceBlock$(b)$
        \STATE $b = b + 1$
        \STATE $\rho = 2$ \COMMENT{Enter power-save mode}
    \ELSE 
        \IF{ValidBlockReceived$(b)$}
            \STATE AppendNewBlock$(b)$
            \STATE $b = b + 1$
            \STATE $\rho = 2$
            \STATE ContinueFindBlockNonce$(b)$
            \IF {$nonce$} 
                \STATE $runnerUp = true$
                \STATE AnnounceRunnerUpBlock$(b)$
            \ELSE 
                \IF{RunnerUpBlockReceived$(b)$}
                    \STATE AbortMining$(b)$ \COMMENT{Enter power-save mode}
                \ENDIF
            \ENDIF
        \ENDIF
    \ENDIF
    \ELSE 
    \STATE \COMMENT{second round: $\rho =2$}
        \IF {$runnerUp$}
            \STATE $nonce = $FindBlockNonce$(b)$ 
            \IF {$nonce$} 
                \STATE AppendNewBlock$(b)$
                \STATE AnnounceBlock$(b)$
            \ELSE 
                \IF{ValidBlockReceived$(b)$}
                \STATE AppendNewBlock$(b)$
                \STATE AbortMining$(b)$
                \ENDIF
            \ENDIF
            \STATE $b = b + 1$
            \STATE $\rho = 1$
            \STATE $runnerUp = false$
        \ELSE
            \STATE \COMMENT{not a runner-up}
            \IF{ValidBlockReceived$(b)$}
            \STATE AppendNewBlock$(b)$
            \STATE $b = b + 1$
            \STATE $\rho = 1$ \COMMENT{Exit power-save mode}
            \ENDIF
        \ENDIF
    \ENDIF
  \ENDLOOP
 \end{algorithmic}
\end{algorithm}

\subsection{Runner(s)-up Election}
In the original PoW, when a puzzle-related block is solved by some miner, all the other network nodes desist the mining of that block and immediately start mining the next block. In Green-PoW, if a valid block is found and the first place winner is elected, the race will continue between miners to also determine the runner-up, i.e., the node that has the second place in the same block race. We denote by $\omega_i$ the first place winner, and $r_i$ the runner-up. Such runner-up ($r_i$) will be the only eligible node to mine in $\rho_{i}^2$, and all the other nodes enter in \textit{mining-save mode} until the end of $\rho_{i}^2$.  As illustrated in \figurename\ \ref{fig:modified_bitcoin}, when a miner $m_j$ receives a block from the first place winner it continues the mining of the same block. Miner $m_j$ either (1) finds the nonce before receiving a block from another miner claiming the second place; in this case node $m_j$ will broadcast an announcement to the entire network that it is the runner-up, and then immediately starts the mining of the next block, or (2) receives a block from another node $m_k$ and subsequently, adds it to its runners-up list ($\mathcal{M}_i^2$) and then enters the \textit{mining-save mode} for one round. $\mathcal{M}_i^2$ serves in the second round to make sure that any received block is originated from a valid runner-up.

Because of the distributed and asynchronous nature of the network, it is possible to have multiple nodes that consider themselves as runners-up when they find the block at nearly the same time. A similar situation can happen for the first place winner and leads to network fork that is solved later by the longest chain rule \cite{forks1}. For the runner-up election, this situation will not cause any problem as any node that successfully mines the block in the first round will be considered as a potential runner-up and can participate in the mining race during $\rho_i^2$. In this case, the set of multiple runners-up $\mathcal{M}_i^2= \{r_i^1, r_i^2, ..., r_i^l\}$. It is worth noting that having multiple nodes as second-place winners does not affect the system inconsistency, but on the contrary, it improves system \textit{liveness} as it increases the chance that a block gets mined during the second round.

\figurename ~\ref{fig:modified_bitcoin} illustrates a simplified case where a miner will automatically switch to \textit{mining-save mode} if it receives a block from another node claiming the second place. Algorithm \ref{alg:greenPoW} also summarizes the different steps of Green-PoW consensus algorithm executed by nodes during mining. As illustrated in \figurename ~\ref{fig:modified_bitcoin} and Algorithm \ref{alg:greenPoW}, a miner can be in one of the following four mining states; 1) mining a block to win the first place in $\rho^1$, (2)  continuing mining a block to win the second place in $\rho^1$, (3) mining a block to win the first place in $\rho^2$, or (4) mining-save mode.

In order to engage a sufficient number of participants in $\rho_i^2$ and improve the system \textit{liveness}, in Green-PoW, even if one node has already claimed to be a runner-up, the other nodes can still continue mining. The decision of a node to continue mining the first round's block depends on the probability of winning the race in $\rho_i^2$. A miner that solves the first round's block very late, will have a very small chance to win as the others have started the mining earlier. For the sake of simplicity, when a node receives the first announcement of a runner-up, it continues mining for only a very short period of time $\eta$ with the hope of quickly finding the nonce and joining $\rho_i^2$. The value of $\eta$ is subject to liveness and energy trade-off and is expected to be determined based on the rate of block generation in the network. In the next section, we will introduce another liveness parameter and will elaborate more on how $\eta$ could be tuned.

\subsection{Second Round Timeout}
\usetikzlibrary{decorations.text}

\begin{figure*}
\centering
\tikzstyle{every state}=[draw=blue!60, fill=blue!5, thick, text width=1.3cm, align=center]
\scalebox{0.8}
{
    \begin{tikzpicture}[->,>=stealth',shorten >=1pt,auto,node distance=2.8cm, semithick]
            \node[state, initial] (A)            {Mining new Block};
            \node[state] (B) [above right=1.cm and 2.5cm of A] {Append Block};
            \begin{pgfonlayer} {fg}
                \node[state,draw=green!60, fill=green!5, fill opacity=0.80,text opacity=1] (C) [above right=0.5cm and 9cm of A]   {mining-save mode};
            \end{pgfonlayer}{fg}
            \draw [->,thick, postaction={decorate,decoration={text along path,text align=center, raise=0.1cm, text={Nonce found}}}]      (A) to [bend right=35] (B);
            \draw [->, thick, postaction={decorate,decoration={text along path,text align=left, raise=0.1cm, text={Announce block}}}]      (B) to [bend left=35] (C);
            \draw [<-,thick, above, postaction={decorate, decoration={text along path,raise=0.1cm,text align=center, text={Block recv. ({$2^{nd}$} round)} }}]      (A) to [bend left=50] (C.north west);
            
            \node[state] (D) [below right=1.cm  and 2.5cm of A] {Continue Mining};
            \draw [->,thick, postaction={decorate,decoration={text along path,text align=center,raise=0.1cm, text={Block recv.}}}]      (A) to [bend right=35] (D);
            
            \node[state] (G) [right=5cm of A] {Update $\mathcal{M}^{2}$ List};
            \draw [->,thick, postaction={decorate,decoration={text along path,text align=center,raise=0.1cm, text={Abort mining}}}]      (G) to [bend left=35] (C);
            
            \draw [->,thick, postaction={decorate,decoration={text along path,text align=center,raise=0.1cm, text={Block recv.}}}]      (D) to [bend right=35] (G);
            
            \node[state] (E) [below =2.3cm of G] {End $1^{st}$ Round};
            \draw [->,thick, postaction={decorate,decoration={text along path,text align=center,raise=0.1cm, text={Nonce found}}}]      (D) to [bend right=35] (E);
            
            \node[state,draw=gray!90, fill=gray!10] (H) [right =2.5cm  of E] {Mining next Block};
            \node[state,draw=gray!90, fill=gray!10] (I) [above right=3cm and 2cm of H] {Append Block};
            \node[state,draw=gray!90, fill=gray!10] (J) [right =2cm of H] {Is in $\mathcal{M}^{2}$ List?};
            \node[state,draw=gray!90, fill=gray!10] (K) [above right=4cm and 4cm of H] {Accept Block};
            \draw [->,thick, postaction={decorate,decoration={text along path,text align=center,raise=0.1cm, text={Announce block }}}]   (E) to (H);
            \draw [->,thick, postaction={decorate,decoration={text along path,text align=center,raise=0.1cm, text={Block recv.}}}]      (H) to [bend right=35] (J);
            \draw [<-,thick, postaction={decorate,decoration={text along path,text align=center,raise=0.1cm, text={No}}}]      (H) to [bend left=35] (J);
            \draw [->,thick, postaction={decorate,decoration={text along path,text align=center,raise=0.1cm, text={Yes}}}]      (J) to [bend right=35] (K);
            \draw [->,thick, postaction={decorate,decoration={text along path,text align=center,raise=0.1cm, text={Nonce found}}}]      (H) to [bend left=35] (I);
            \draw [<-,thick, postaction={decorate,decoration={text along path,text align={left indent = {.8\dimexpr\pgfdecoratedpathlength\relax}},raise=0.1cm, text={Announce block}}}]      (A) to [bend left=70] (I);
            \draw [<-,thick, postaction={decorate,decoration={text along path,text align={left indent = {.8\dimexpr\pgfdecoratedpathlength\relax}},raise=0.1cm, text={Abort mining}}}]      (A.north) to [bend left=60] (K);
            
            \node[] (v1) [above left=3.2cm and 1cm  of C] {};
            \node[] (v2) [below left=7cm  and 0.1cm of C] {};
            \draw [-, dashed, red, thick]      (v1) to [bend left=30] (v2);
            \node[blue] (r1) [left=0.5cm  of v2] {First round ($\rho^1$)};
            \node[] (r2) [right=0.5cm  of v2] {Second round ($\rho^2$)};
    \end{tikzpicture}
    }
\caption{The state diagram of a miner in Green-PoW.}
\label{fig:modified_bitcoin}
\end{figure*}
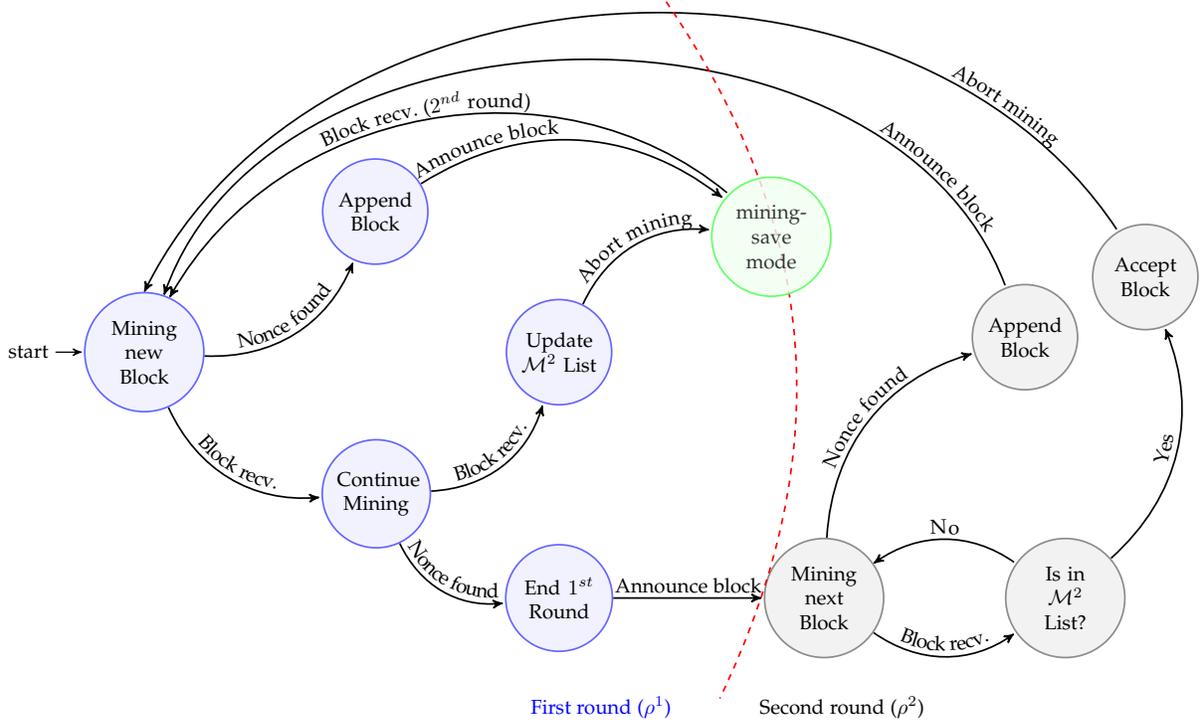    
Since only nodes that successfully mine the first round block may be part of $\mathcal{M}_i^2$, the number of potential miners in $\rho_i^2$, i.e., $|\mathcal{M}_i^2|$, is naturally limited. While this is advantageous from an energy conservation point of view, it is possible that the system goes to a deadlock and the next block does not get generated. This can happen for instance if miners in $\mathcal{M}_i^2$  are isolated from the rest of the network, inadvertent in case of network segmentation, or intentionally by an adversary who launches an eclipse attack \cite{eclipse}. To mitigate this problem and ensure system \textit{liveness}, Green-PoW employs a time-out at the beginning of each $\rho_i^2$. The time-out should be greater than the average time needed to generate a new block. If a block announcement is not received before the time-out, inactive nodes quit the \textit{mining-save mode} and start immediately $\rho_i^2$. However, because of the asynchronous nature of the network and the malicious behavior of some nodes, the introduction of time-out may lead to some special cases that we discuss in the following:
\begin{itemize}
    \item A malicious node that is not eligible to participate in $\rho_i^2$, may try to start the mining before the time-out in order to get an advantage (receive a reward) over other non-participating nodes. However, this is very risky for the attacker as the success of such manipulation depends on the probability that no block will be received from eligible nodes ($\mathcal{M}_i^2$); in such a case the attacker could be wasting a lot of energy if it does not win.  
    \item Due to network asynchronicity, it is possible that some nodes will receive a valid block from an honest ineligible participant while their time-out is not yet ended. In this situation, these nodes will initially reject the block, but because the majority will accept it, the block will appear in the longest chain and the minority will end up accepting it.
\end{itemize}
It is worth noting that both the time-out and $\eta$ help in defining a good balance between system liveness and energy-efficiency. Growing $\eta$ improves system liveness by increasing $|\mathcal{M}_i^2|$, yet it diminishes energy-efficiency as more miners will participate in $\rho_i^2$. By using the time-out, Green-PoW can ensure liveness, but it may also increase the energy consumption if   $|\mathcal{M}_i^2|$ is very small, i.e., $\eta$ is very short, as timing-out $\rho_i^2$ is likely to happen frequently. Therefore, a careful selection of $\eta$ is crucial for Green-PoW to ensure the desired energy-efficiency. More hints about the typical choice of the  time-out and $\eta$ parameters are given in Section \ref{sec:performance}.

\subsection{Second Round Mining Difficulty}
As presented in Section \ref{sec:mining-difficulty}, given that the total mining power of the network can change over time, the mining difficulty is dynamically adjusted to ensure that blocks are generated at a nearly constant rate. In Green-PoW, because the total hash power decreases drastically in each second round mining, compared to the first round, a new difficulty level defined specifically for the second round is required. Let $D^1$ and $D^2$ denote the difficulty level to consider during $\rho^1$ and $\rho^2$, respectively. A block is considered as valid only if its hash value respects the target hash (e.g., in the block header), which is calculated based on the difficulty level of the corresponding mining round. The difficulty level is initially set to the minimum value and updated every $T$ period of time (every 2 weeks in Bitcoin). Using equation (\ref{eq:1}) and (\ref{eq:2}) presented in Section \ref{sec:mining-difficulty}, the new difficulties $D_{j}^1$ and $D_{j}^2$ for the $j^{th}$ $T_j$ period can be calculated as follow:
\begin{equation}
    D_j^1 = D_{j-1}^1 \left(\frac{T_{E}}{T_{Avg^1}}\right)
    \label{eq:diff:first_round}
\end{equation}

\begin{equation}
    D_j^2 = D_{j-1}^2 \left(\frac{T_{E}}{T_{Avg^2}}\right)
    \label{eq:diff:second_round}
\end{equation}

Where $T_{E}$ is the expected average block time (10 minutes in Bitcoin), and $T_{Avg^1}$ and $T_{Avg^2}$ are the actual average block-time calculated over the last $T_j$, for blocks generated in $\rho^1$ and $\rho^2$, respectively. Note that because it may happen that some blocks in the second round are generated with the participation of all network miners, in case of timing out, these blocks will be considered for the adjustment of $D^1$ but not $D^2$. To distinguish between blocks generated using a small subset of miners from those using all network miners, the target value in the header of each  block will be verified.

The mining in $\rho^2$ is similar to the mining in the original PoW with the exception that only miners in the corresponding $\mathcal{M}^2$ can participate, and use the second mining difficulty $D^2$. When a valid block is formed and propagated to the entire network, as shown in \figurename~\ref{fig:modified_bitcoin}, the other active miners in $\mathcal{M}^2$ stop the mining of the current block and start new mining epoch. The passive miners ($\mathcal{M} \setminus \mathcal{M}^2$) also leave their \textit{mining-save mode} and join the others in a new epoch. 

\section{Security Analysis}
\label{sec:security_analysis}
\subsection{Transaction Censorship}

The potential of transaction censorship exists in public blockchains because of the limited size of the block and the asynchronous nature of the network, which makes it difficult to verify which set of transactions a particular block must include \cite{bitcoin-ng,censorship}. Usually, miners select only a subset of transactions from their pool of pending transactions to not exceed the block-size limit and tend to prioritize transactions with higher fees to maximize their profit.  Such freedom on selecting transactions to include in a block gives adversary miners the opportunity to censor some transactions from being added to the next block, even those with high fees.

By design, PoW extenuates this concern because the censorship time $t_c$ is bounded by the average block generation time $1/\lambda$, and restricted by the fact that a malicious miner $m_c$ must be the winner. However, a powerful attacker that may successfully mine and win $k$ consecutive blocks will delay the inclusion of some urgent transactions for a longer time, i.e., $t_c = k/\lambda$.
In Green-PoW, a malicious miner $m_c$ can still censor transactions for $t_c$ during $\rho^1$, yet because the winner miner in $\rho^1$ cannot participate in $\rho^2$, the effect of such an attack is limited to only one mining round. As a result, Green-PoW can reduce the censorship time to nearly $50\%$, which guarantees users better transaction time and reduces the intensity of a potential denial of service attack launched by powerful miners. The only case when a winner in $\rho^1$ can also participate and win in $\rho^2$, is after timing out $\rho^2$. However, this can only happen under specific conditions, such as separating the set $\mathcal{M}^2$ from the rest of the network, which is difficult to be controlled by an attacker. 

\subsection{Mining Centralization}
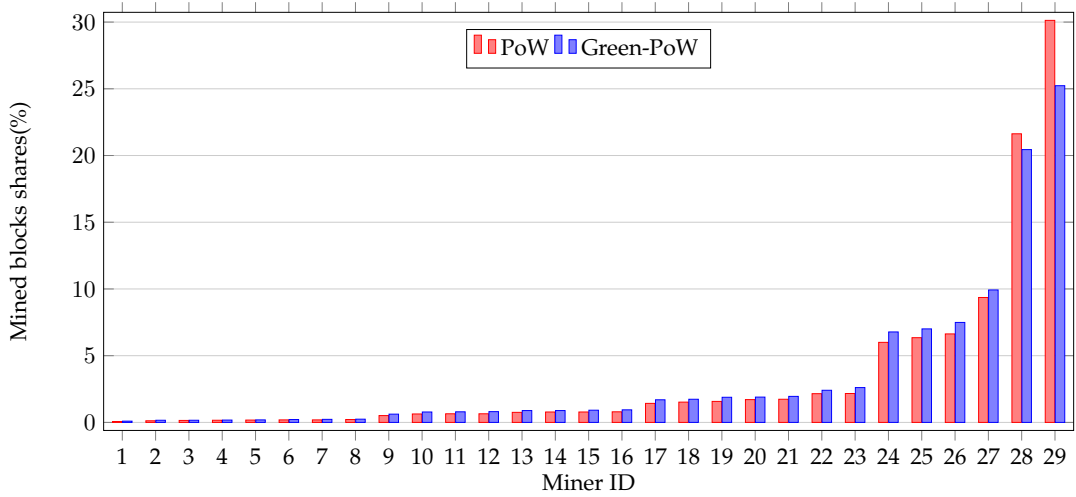
\begin{figure*}
\centering
\begin{center}
\scalebox{0.9}{
\begin{tikzpicture}
\begin{axis}[
      width=9cm,
      ybar=0.2pt,
      ymin=0,
      x=14,
      enlargelimits=0.02,
      ymajorgrids=true, grid style={gray!40}, 
     legend style={at={(0.5,.96)},anchor=north,legend columns=-1}, 
      xlabel={Miner ID}, ylabel={Mined blocks shares(\%)},
      symbolic x coords={1,2,3,4,5,6,7,8,9,10,11, 12, 13,14,15,16,17,18,19,20,21,22,23,24,25,26,27,28,29},
      xtick={1,2,3,4,5,6,7,8,9,10,11, 12, 13,14,15,16,17,18,19,20,21,22,23,24,25,26,27,28,29},
      bar width=4,
    ]
    
\addplot [fill=red!50, draw=red, fill opacity=1] 
coordinates { 
(1, 0.06724949562878278)
(2, 0.121049092131809)
(3, 0.14794889038332212)
(4, 0.16139878950907868)
(5, 0.17484868863483524)
(6, 0.1882985877605918)
(7, 0.1882985877605918) 
(8, 0.21519838601210492)
(9, 0.5110961667787491)
(10, 0.6321452589105582)
(11, 0.6455951580363147)
(12, 0.6455951580363147)
(13, 0.7531943510423672)
(14, 0.7800941492938803)
(15, 0.7800941492938803)
(16, 0.7935440484196369)
(17, 1.425689307330195)
(18, 1.5198386012104907)
(19, 1.5736381977135172) 
(20, 1.7081371889710828) 
(21, 1.735036987222596) 
(22, 2.151983860121049) 
(23, 2.1654337592468056) 
(24, 5.9986550100874245) 
(25, 6.348352387357095) 
(26, 6.630800268997983) 
(27, 9.361129791526563) 
(28, 21.627437794216544) 
(29, 30.12777404169469) 
};

\addplot [fill=blue!50, draw=blue, fill opacity=1]
coordinates { 
(1, 0.0941492938802959)
(2, 0.16139878950907868) 
(3, 0.16139878950907868) 
(4, 0.17484868863483524) 
(5, 0.1882985877605918) 
(6, 0.21519838601210492) 
(7, 0.22864828513786145) 
(8, 0.242098184263618) 
(9, 0.6186953597848016) 
(10, 0.7800941492938803) 
(11, 0.7935440484196369) 
(12, 0.8069939475453934) 
(13, 0.8876933422999328) 
(14, 0.8876933422999328) 
(15, 0.9145931405514458) 
(16, 0.941492938802959) 
(17, 1.6946872898453262) 
(18, 1.735036987222596) 
(19, 1.882985877605918) 
(20, 1.8964357767316746) 
(21, 1.9502353732347006) 
(22, 2.4075319435104237) 
(23, 2.609280430396772) 
(24, 6.778749159381305) 
(25, 7.007397444519166) 
(26, 7.491593813046402) 
(27, 9.926025554808339) 
(28, 20.443846671149966) 
(29, 25.232010759919298)};

 \legend{PoW, Green-PoW}
\end{axis}
\end{tikzpicture}
}
\caption{Shares of miners in Ethereum Vs. Green-PoW based Ethereum, calculated using real data of the latest 7438 blocks imported from the main Ethereum network.}
\label{fig:shares_comp}
\end{center}
\end{figure*}
In PoW-based networks, such as Bitcoin or Ethereum, the mining power is concentrated among a relatively small number of miners (pools) which makes the crypto-system highly susceptible to censorship or even $51\%$ attacks \cite{51attack}. Less powerful miners are usually unfortunate to generate a new valid block in the presence of other superior miners. This will eventually lead to a monopoly based system, where a small percentage of the network earns the highest rewarding shares. By design, Green-PoW can reduce the monopoly of powerful miners, since generating consecutive blocks by the same miner is likely not possible. A miner that wins the mining race in $\rho^1$ is not allowed to participate in $\rho^2$, and consequently gives the other nodes the chance to win with less competition. The only case when it is still possible that a winner in $\rho^1$ also wins in $\rho^2$, is after timing out $\rho^2$.

To better illustrate the impact of Green-PoW on mining shares distribution, we have imported the latest 7437 blocks (50 days) from the Ethereum main network. We plot in \figurename~\ref{fig:shares_comp} the corresponding shares of each miner as the ratio of its mined blocks relative to the total block count, i.e., 7437. In the same plot, we also include the corresponding shares of each miner when applying Green-PoW to the same blocks. In Green-PoW, two, three, four, or more consecutive blocks are highly improbable to be mined by the same miner. It is worth mentioning that two consecutive blocks could be generated by the same miner under a special condition, where a miner wins the second round block and the next block in the first round of the following epoch. For this reason, we subtracted one block from only 50\% of two consecutive blocks cases. As shown in \figurename~\ref{fig:shares_comp}, the corresponding shares of the most powerful miners, i.e., miner numbers 28 and 29, in Green-PoW are reduced, compared to the case of the original PoW. Such an impact limits the dominance that the most powerful miners may have on the network. For example, the most powerful miner had a $5\%$ reduction in its share, which is redistributed among other nodes. The reduction can be more significant if the computing power varies widely among the nodes. Subsequently, better share distribution between miners could be achieved.

\subsection{Fork Occurrences}
\label{sec:fork:occurences}
As discussed in Section \ref{sec:fork}, a fork in blockchain can happen when multiple miners find a block almost simultaneously. More generally, a fork occurs whenever a miner $m_j$ finds a block while another miner $m_i $ has already formed a valid block without being aware of it. This situation is likely to occur in a large network having a long propagation delay. 

To illustrate the effect of propagation delay on the fork occurrences, Decker and Wattenhofer \cite{decker2013information} have conducted a theoretical study and presented an approximate model to predict the rate at which forks can occur. For a newly found block $b_i$ by $m_i$, the probability of fork occurrence, i.e., conflicting blocks will be found by other miners before being aware of $b_i$, is estimated by (i) determining the number of unaware miners at time $t$, and (ii) the probability that each unaware miner will find a conflicting block during that time. Given a ratio of unaware miners $u(t)$ about $b_i$ during time $t$, the probability of having a fork ($F$) on the network can be expressed as follow:    

\begin{equation}
    Pr[F > 0] = 1-(1-P_b)^{\int_0^\infty u(t) \ dt }
\end{equation}
Where $P_b$ is the probability of a block being found by the network at a given time $t$. From the formula, it is clear that the fork rate is proportional to the ratio of unaware miners and thus proportional to the total number of miners in the network.

In Green-PoW, during the second round, the ratio of active miners is very small compared to the first round. Subsequently, the ratio of unaware miners $u$ is also very small.

\section{Performance Evaluation}
\label{sec:performance}
In this section, we evaluate the power consumption of Green-PoW  using stochastic analysis. As modeled in the Bitcoin white paper \cite{bitcoin} and in \cite{kraft2016difficulty, fullmer2018analysis, bowden2018block}, the Bitcoin mining process could be well-approximated as a Poisson process with a deterministic rate $\lambda$ which represents the mining rate or the \textit{average time} between block-arrival events. In the following, we first formulate the mining process in both PoW and Green-PoW as a Poisson process and then assess the power saving achieved by Green-PoW. 

\subsection{Average Power Saving in Green-PoW}
Let $\mathcal{M}$ denote the set of $n$ miners in the network $\mathcal M=\{m_1, m_2, ..., m_n\}$. Each miner $m_i \in \mathcal M$ has a fraction $h_i$ of the total hashing power in the network $\mathcal{P}$, so that it mines a new block at a rate of $h_i\lambda$, where $\sum h_i =1$. As explained in Section \ref{sec:mining-difficulty}, the difficulty of finding a block is dynamically adjusted to ensure that a block is generated every $1/\lambda$ seconds in expectation with a rate $\lambda$ ($\lambda = 1/600$ in Bitcoin). The inter-arrival times of consecutive blocks follow Exponential distribution with the same rate parameter  $\lambda$, whose cumulative distribution function is:
\begin{equation}
    \mathrm{Pr}[T \leq t] = 1- e^{-\lambda t}
\label{eq:exp}
\end{equation}
In PoW, each miner $m_i$ spends on average $1/\lambda$ and consumes energy $\mathrm{E}_i$ which is proportional to its hashing power $h_i$ and can be expressed as:
\begin{equation}
\mathrm{E}_i = \frac{1}{\lambda} \ h_i \ \mathcal P
\label{eq:powerPoWminer}
\end{equation}
Therefore, the average total energy $\mathrm{E}$ consumed by the network to generate a block is inversely proportional to the block generation rate $\lambda$:
\begin{equation}
\mathrm{E} = \mathcal P \frac{1}{\lambda}\ \sum_{i=1}^{n} h_i  = \mathcal P \frac{1}{\lambda} 
\label{eq:powerPoW}
\end{equation}
In Green-PoW, a block is either generated during the first or the second round. In the first round, compared to PoW mining, additional energy is consumed in order to select the second-place winners. This additional energy depends on the number of second-place winners and the time they need to complete the mining and form a valid block. Assuming that $m_f$ is the first winner, $m_s, m_{s+1}, ..., m_{s+k}$ are $k$ runners-up, and $t_s, t_{s+1}, ..., t_{s+k}$ the respective time needed by each of the runner-up to find the block. Thus the average total energy $\mathrm{E_{1^{st}}}$ consumed by the network during the first round can be expressed as follow:
\begin{gather}
\mathrm{E_{1^{st}}} = \mathcal P \left( \frac{1}{\lambda} + \sum_{i=s}^{k} t_i (1-h_f +\sum_{j=i-1}^{k-1}h_j)\right),  
\label{eq:power1round}
\end{gather}
\hspace{2cm} where  $h_{s-1} = 0$ \vspace{0.1cm} \\ 
In the second round, the average consumed energy is proportional to the time needed to generate a block ($1/\lambda$) and the total hashing power of the runners-up. For simplicity, we do not consider the scenario where the set of runners-up fail to generate a block, and other miners start the mining process after the timeout:

\begin{equation}
\mathrm{E_{2^{nd}}} = \mathcal P \frac{1}{\lambda} \sum_{i=s}^{k} h_i 
\label{eq:power2round} 
\end{equation}
From equations (\ref{eq:powerPoW}), (\ref{eq:power1round}) and (\ref{eq:power2round}), the power saving in Green-PoW can be, therefore, expressed as follow:

\begin{equation}
\mathrm{E_{save}} =  2\ \mathrm{E} - (\mathrm{E_{1^{st}}} + \mathrm{E_{2^{nd}}})
\label{eq:powerSave}
\end{equation}

\subsection{Experimental Setup}
In order to determine the time needed to select $k$ runners-up and thus, calculate the energy spent in the first and the second round, we basically used the inverse function of the CDF in equation (\ref{eq:exp}) and feed it different probability values $p$ from a Uniform$(0,1)$ distribution to generate the blocks inter-arrival times $t_{}$:
\begin{equation}
t =  -\frac{1}{\lambda}\log{(1-p)}
\label{eq:invExp}
\end{equation}
The time when a runner-up finds a valid block, during the first round, can be estimated as follow:
\begin{equation}
t =  -\frac{1}{\lambda(1-h_{prev})}\log{(1-p)}
\label{eq:invExp}
\end{equation}
where $h_{prev}$ is the sum of the hashing power of all its predecessor runners-up including the first winner of the round. The time $t$ is increasing for every newer runner-up as the ratio of the total network power is decreasing ($1-h_{prev}$).

We conduct extensive simulation and average the power saving in Green-PoW over $100,000$ blocks. We consider three network sizes with 100, 200, and 300 miners, and different hashing power using Uniform and Normal distribution. Table \ref{tab:param} summarises the simulation parameters used to assess power saving.

\begin{table}[h]
\centering
\scalebox{0.95}{
\begin{tabular}{|l|l|}
\hline
\textbf{Parameter}         & \textbf{Value}      \\ \hline
\# blocks                  & $100,000$              \\\hline
\# miners                  & $[100, 200, 300]$ \\\hline
\# second winners          & $[1, 2, ..., 10]$ \\\hline
Hashing power dist. & $\%$ of miners having $50\%$ of total hash power 
\\ & $[2\%, 5\%,10\%,20\%, 50\%]$\\ 
\hline
\end{tabular}
}
\caption{Simulation Parameters}
\label{tab:param}
\end{table}

\subsection{Results}
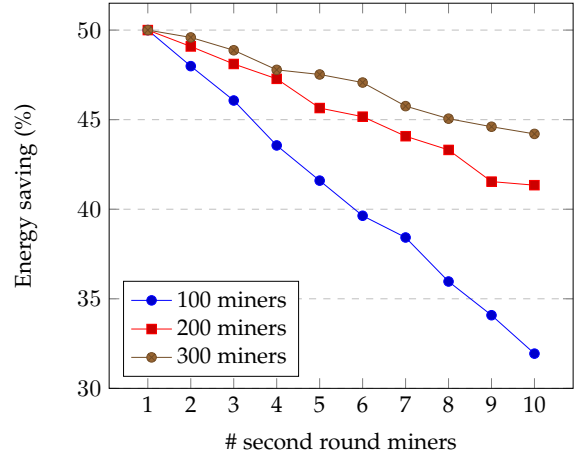
\begin{figure}
\centering
\scalebox{0.9}{
\begin{tikzpicture}
\begin{axis}[xlabel={\# second round miners}, ylabel={Energy saving (\%)},
     ymin=30, ymax=51.5, ytick={30,35,40,45,50,55},
    xtick={1, 2, 3, 4, 5, 6, 7, 8, 9, 10}, xticklabels={1, 2, 3, 4, 5, 6, 7, 8, 9, 10}, legend columns=1 ,legend pos=south west, ymajorgrids=true, grid style=dashed,]
    
    \addplot coordinates { (1, 50.0) (2, 47.98742038709414) (3, 46.069450359167803) (4, 43.560816482814971) (5, 41.596774871020649) (6, 39.63594609558843) (7, 38.42218295180426) (8, 35.965044664276064) (9,34.082630675370723) (10, 31.939095166230985) };

    \addplot coordinates { (1, 50.0) (2, 49.084499896492197) (3, 48.108072735481962) (4, 47.278366294025005) (5, 45.648096021980244) (6, 45.167540078945869) (7, 44.075507722952189) (8, 43.310342149348337) (9, 41.542417999732152) (10, 41.341812744785543) };

    \addplot coordinates { (1, 50.0) (2, 49.590302563211409) (3, 48.873074749029357) (4, 47.780456619418707) (5, 47.522089628978932) (6, 47.071703657527102) (7, 45.755866435412877) (8, 45.054943728345876) (9, 44.605129789555498) (10, 44.204690072409655) };

\legend{100 miners, 200 miners, 300 miners,400 miners, 500 miners }
\end{axis}
\end{tikzpicture}
}
\caption{Energy saving ratio Vs. number of second round miners} 
\label{fig:power_saving:miners}
\end{figure}

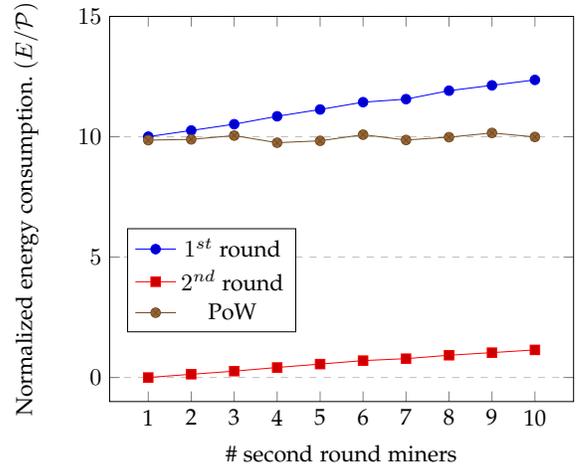
\begin{figure}
\centering
\scalebox{0.9}{
\begin{tikzpicture}
\begin{axis}[xlabel={\# second round miners}, ylabel={Normalized energy consumption. $(E/\mathcal P)$},
ymin=-1, ymax=15, ytick={0,5,10,15}, xtick={1, 2, 3, 4, 5, 6, 7, 8, 9, 10}, xticklabels={1, 2, 3, 4, 5, 6, 7, 8, 9, 10},legend columns=1 ,legend style={legend style={at={(0.22,.45)},anchor=north}}, ymajorgrids=true, grid style=dashed,]
    
\addplot coordinates { (1, 10.005904225288964) (2, 10.261325637826515) (3, 10.522097891179367) (4, 10.850812187131309) (5, 11.133019484231708) (6, 11.437213870458356) (7, 11.562555887902493) (8, 11.91440342448926) (9, 12.135615937374544) (10, 12.361731623860464) };

\addplot coordinates { (1, 0) (2, 0.13388774915414134) (3, 0.26214343660344808) (4, 0.41646439236626176) (5, 0.556432269729771) (6, 0.70109565936980678) (7, 0.78230559005486378) (8, 0.92421220690930106) (9, 1.0324940422574624) (10, 1.1458045535826198) };

\addplot coordinates { (1, 9.8592738438349095) (2, 9.8917638463545554) (3, 10.049782578986907) (4, 9.752672659115369) (5, 9.8316469853426511) (6, 10.085954643580893) (7, 9.8661036755847853) (8, 9.9825764709926794) (9, 10.158868262270349) (10, 9.9925599596075152) };

\legend{$1^{st}$ round, $2^{nd}$ round, PoW}
\end{axis}
\end{tikzpicture}
}
\caption{ Normalized energy consumption Vs. number of second round miners
}
\label{fig:power_consump}
\end{figure}

\figurename~\ref{fig:power_saving:miners} and \figurename~\ref{fig:power_consump} show the impact of the number of runners-up and the size of the network on the total energy consumption in Green-PoW.  
\figurename~\ref{fig:power_saving:miners} illustrates the ratio of power saving in Green-PoW with respect to the original PoW when varying the number of second round contenders for different network sizes. When only one node mines the block in the second round, the saving power is nearly $50\%$ regardless of the size of the network. However, for a larger number of winners, the saving drop to nearly $32\%$, $41\%$, and $44\%$ for networks of $100$, $200$, and $300$ nodes, respectively. 
We also evaluate the total energy consumption in PoW and Green-PoW during the first and the second round and plot the results in \figurename~\ref{fig:power_consump}. For a network of $100$ miners, as shown in the figure, in Green-PoW the energy consumption during the second round is nearly $10\%$  of that of the first round; such dramatic energy saving is due to the fact that only few nodes are participating in the mining process during the second round. In PoW, the average energy consumption is almost constant and is $8$-$10$ times more than the second round of Green-PoW. Green-PoW consumes more energy than PoW in the first round since the nodes continue mining the same block in order to determine the runners-up. Nonetheless, the average of the first and second rounds is about $30$-$50\%$ less than PoW.

\begin{figure}
\centering
\scalebox{0.9}{
\begin{tikzpicture}
\begin{axis}[xlabel={Hash-power distribution}, ylabel={Energy saving (\%)},
     ymin=10, ymax=51.5, ytick={15,25,35,45,55},
    xtick=data, xticklabels={2\%, 5\%, 10\%, 20\%, 50\%}, ymajorgrids=true, grid style=dashed,]
    
    \addplot coordinates {  (1, 14.72) (2, 33.956591165822658) (3, 41.11) (4, 44.86) (5, 47.41)};

\end{axis}
\end{tikzpicture}
}
\caption{Energy saving ratio Vs. hash-power distribution} 
\label{fig:power_saving:hashrate:dist}
\end{figure}
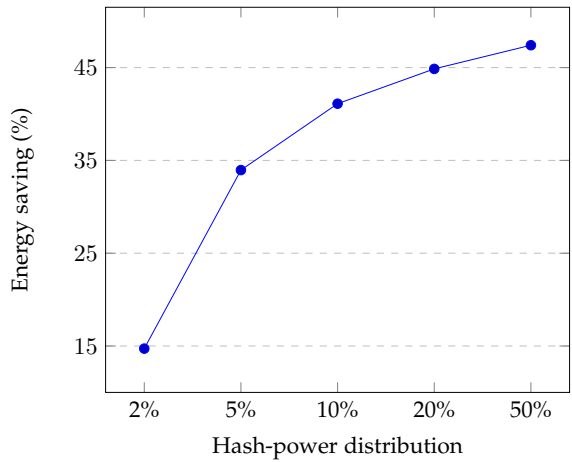

We also assess in \figurename~\ref{fig:power_saving:hashrate:dist} the impact of a different distribution of the hashing power on the energy-saving in Green-PoW. We consider a network of  $200$ miners and engage $5$ nodes to mine in $\rho^2$. We distribute the hashing power among network miners by varying the percentage of miners that hold $50\%$ of the total network hash power, while assuming the remaining power is equally distributed among the other $50\%$ of the network. When $50\%$ of the network, i.e., 100 miners,  equally hold $50\%$ of the hashing power, this means that all miners have exactly the same portion of hashing power ($0.5\%$). As illustrated in the figure, when a small portion of miners ($2\%$) holds most of the hashing power ($50\%$), the energy-saving in Green-PoW is minimal, however, when the power is equally distributed among miners, Green-PoW achieves its maximal saving. This is mainly due to the fact that the energy-saving in Green-PoW depends on the mining power of the second round contenders. When some of them  have high power,  per equation (\ref{eq:diff:second_round}) the mining difficulty will be increased and consequently, more energy needs to be spent in order to find the second round block. Vise versa, when they have small hashing power, less energy will be consumed in order to find the block in the second round as the mining difficulty will be reduced.




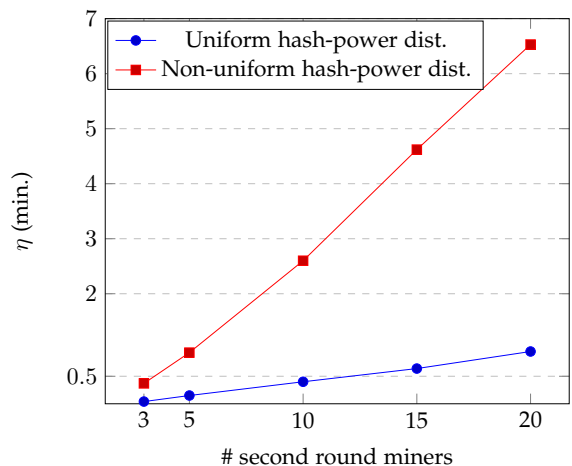
\begin{figure}
\centering
\scalebox{0.9}{
\begin{tikzpicture}
\begin{axis}[xlabel={\# second round miners}, ylabel={  $\eta$ (min.)},
ymin=0, ymax=7, ytick={0.5, 2,3,4,5,6,7}, xtick={3,  5, 10, 15, 20}, xticklabels={ 3,  5, 10, 15, 20},legend columns=1 ,legend style={legend style={at={(0.41,.995)},anchor=north}}, ymajorgrids=true, grid style=dashed,]
    
\addplot coordinates { (3, 0.04) (5, 0.15) (10, 0.40) (15, 0.64) (20, 0.95) };
\addplot coordinates { (3, 0.37) (5, 0.93) (10, 2.60) (15, 4.62) (20, 6.53) };

\legend{Uniform hash-power dist. , Non-uniform hash-power dist.}
\end{axis}
\end{tikzpicture}
}
\caption{ Average time ($\eta$) between the first and last considered runner-up to be include in $M_i^2$.
}
\label{fig:eta}
\end{figure}

\subsection{Time-out and $\eta$ Selection}
As we discussed in Section \ref{sec:green-pow}, the time-out and $\eta$ are two important parameters that help in striking a good balance between system liveness and energy-efficiency. The $\eta$ parameter defines the additional time a particular node needs to spend in mining during $\rho_i^1$ after hearing from the first considered runner-up. This time can be set as a function of the number of miners that we want to have in the second round, i.e., a function of $|\mathcal{M}_i^2|$. To capture the effect of $\eta$, we plot in \figurename~\ref{fig:eta} the time needed in order to have a specific size of $\mathcal{M}_i^2$. We consider the same simulation parameters as before, and we plot the time between the first and last considered runner-up, when having, 3, 5, 10 15, and 20 miners in the second round. We also consider different distributions of the hashing power in the network. In the case of uniformly distributed hash power among miners, the value of $\eta$ does not increase much with the number of second round miners; however, when the distribution is not uniform, specifically, when $50\%$ of the power is held by only $5\%$ of the miners, $\eta$  increases significantly. This is because more time is needed to wait for less-powerful nodes to mine a block and be able to join other miners in $\rho_i^2$.

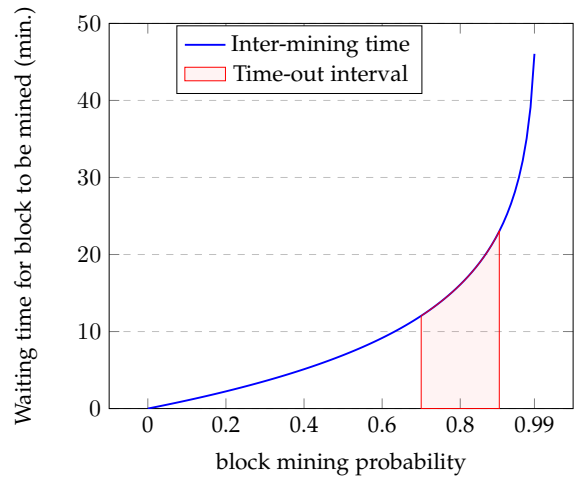
\begin{figure}
\centering
\scalebox{0.9}{
\begin{tikzpicture}
\begin{axis}[xlabel={block mining probability}, ylabel={  Waiting time for block to be mined (min.)},
ymin=0, ymax=50, ytick={0, 10, 20, 30, 40, 50}, xtick={0, 0.2,  0.4,  0.6,  0.8,  0.99}, xticklabels={ 0, 0.2,  0.4,  0.6,  0.8,  0.99},legend columns=1 ,legend style={legend style={at={(0.41,.995)},anchor=north}}, ymajorgrids=true, grid style=dashed,]
    
\addplot[color=blue,domain=0:0.99,samples=100, thick] {-ln(1.0 - x) / (1/10))};
\addlegendentry{Inter-mining time}

   \addplot+[mark=none,
        domain=0.7:0.9,
        samples=100,
        fill=red,
        color= red,
        fill opacity=0.05,
        area legend
        ]{-ln(1.0 - x) / (1/10))} \closedcycle; 
        \addlegendentry{Time-out interval}
\end{axis}
\end{tikzpicture}
}
\caption{ Time to wait for a block to be mined in the second round Vs. the corresponding probability.
}
\label{fig:timeout}
\end{figure}

We also plot in \figurename~\ref{fig:timeout} the required time for a block to be mined in the second round. As discussed previously, the inter block generation (mining) time follows Exponential distribution with the same rate parameter $\lambda$ ($1/600$ in Bitcoin). Using equation (\ref{eq:invExp}) we plot the mining time between two consecutive blocks (time between the first round block and second round block) for different probability. A safe time-out can be chosen as the duration of time ensuring that a block will be mined with a high probability. For instance, for a block to be mined with a probability between [0.7, 0.9]  a network needs to wait for a time between [12, 23] minutes. Therefore, a typical time-out can be chosen from this interval. Note that having different hash power distribution will not affect the block generation time, as the defined difficulty $D_i^2$ in equation (\ref{eq:diff:second_round}) ensures that a block is mined at a constant rate on the average (10 minutes in Bitcoin).


\section{Conclusion}
\label{sec:conclusion}
In this paper, we have proposed a novel and energy-efficient consensus algorithm, called Green-PoW, for a public blockchain. In our algorithm, the overall energy consumption during mining is reduced by up to $50\%$ compared to the original PoW. Green-PoW achieves its goal by taking advantage of the energy spent during one block mining to also elect a small number of miners that will exclusively mine the next block. In Green-PoW, time is divided into epochs that consist of two mining rounds. The first round is similar to mining in the original PoW with the exception that a small additional power is spent in order to qualify a subset of miners to exclusively contend in the second round. In the second round, where most of the mining power is saved, only the elected miners during the previous round have the right to participate and compete for forming a new block. To validate the performance of Green-PoW, extensive simulations have been conducted to mainly assess the energy saving compared to the original PoW. The results demonstrated the efficiency of the solution where up to $50\%$ of the mining energy can be saved for a large network with equally distributed hashing power. We also have studied key security properties and shown the advantage of Green-PoW in reducing fork occurrences, the effect of censorship attack, and mining centralization.

\bibliographystyle{IEEEtran}
\bibliography{references}

\end{document}